\documentclass{aa}
\usepackage{graphicx}
\usepackage[varg]{txfonts}
\usepackage{longtable}
\usepackage{lscape}

\begin{document}

\title{Activity time series of old stars from late F to early K. II. Radial velocity jitter and exoplanet detectability}

\titlerunning{Activity time series of old stars from late F to early K. II.}

\author{N. Meunier \inst{1}, A.-M. Lagrange \inst{1} 
  }
\authorrunning{Meunier et al.}

\institute{
Univ. Grenoble Alpes, CNRS, IPAG, F-38000 Grenoble, France\\
\email{nadege.meunier@univ-grenoble-alpes.fr}
     }

\offprints{N. Meunier}

\date{Received ; Accepted}

\abstract{The  effect of stellar activity on radial velocity (RV) measurements appears to be a limiting factor in detecting Earth-mass planets in the habitable zone of a star that is similar to the Sun in spectral type and activity level. It is crucial to estimate whether this conclusion remain true for other stars with  current correction methods. }
{We built realistic time series in radial velocity and chromospheric emission for old  main-sequence F6-K4 stars. We studied the effect of the stellar parameters we investigate on exoplanet detectability. The stellar parameters are spectral type, activity level, rotation period, cycle period and amplitude, latitude coverage, and spot constrast, which we chose to be in ranges that are compatible with our current knowledge of stellar activity.}
{This very large set of synthetic time series allowed us to study the effect of the parameters on the RV jitter and how the different contributions to the RV are affected in this first analysis of the data set. The RV jitter was used to provide a first-order detection limit for each time series and different temporal samplings. }
{We find that the coverage in latitude of the activity pattern and the cycle amplitudes have a strong effect on the RV jitter, as has stellar inclination. RV jitter trends with B-V and LogR'$_{\rm HK}$ are similar to observations, but activity cannot be responsible for RV jitter larger than 2-3 m/s for very quiet stars: this observed jitter is therefore likely to be due to other causes (instrumental noise or stellar or planetary companions, e.g.). Finally, we show that based on the RV jitter that is associated with each time series and using a simple  criterion, a planet with one Earth mass and a period of one to two years probably cannot be detected with current analysis techniques, except for the lower mass stars in our sample, but very many observations would be required. The effect of inclination is critical. }
{The results are very important in the context of future RV follow-ups of transit detections of such planets. We conclude that a significant improvement of analysis techniques and/or observing strategies must be made to reach such low detection limits. }


\maketitle

\section{Introduction}

In the past few years, we undertook several steps to be able to produce realistic radial velocity (hereafter RV) time series for stars other than the Sun. After modeling the solar RV using observed magnetic structures \cite[][]{lagrange10b,meunier10a}, we developed a complex model in which we randomly generated the structures using our current knowledge of the Sun \cite[][]{borgniet15}. This model allowed us to simulate complex patterns of activity for the Sun viewed from any point of view. We also computed other variables, such as brightness variations and astrometry.  
 Finally, we adapted the model to other stars  (different activity levels and spectral types) that we described in \cite{meunier19}, hereafter Paper I. 
Other models considering complex activity patterns have been proposed by \cite{herrero16} to reproduce the contributions of spots and plages and by \cite{santos15} for spots alone. The simulations made in \cite{dumusque16} include all effects for a few configurations.

In this paper we perform a first analysis based on the RV jitter. Although we know that using RV jitter remains a simple approach, it is widely used in the literature: simple models of activity \cite[][]{saar97,hatzes02,saar03,wright05}  have been used to predict RV jitter from other data \cite[e.g.,][]{martinez10,arriagada11}.  
Therefore it is interesting to propose a first application of the RV jitter derived from the models to explore the limits of exoplanet detectability as a function of spectral type and activity level beyond the solar case. We consider a threshold in RV jitter that corresponds to the current status of correction methods derived from \cite{dumusque16} and \cite{dumusque17} and apply it to derive an approximate  detection limit as a function of spectral type and activity level. 

The outline of the paper is the following. In Sect.~2 we recall the model principles, parameters, and the observables, which are produced for each set of parameters from Paper I. Sect.~3 describes the effect of the parameters on the RV jitter. We also studied the individual components due to magnetic activity of the whole RV signal and the effect of several parameters  on the corresponding RV jitter. In Sect.~4 we compare the RV jitter we obtained in our simulations with observations, and we investigate the RV jitter at different scales. In Sect.~5 we propose a first application of these RV jitter to exoplanet detectability, and we conclude in Sect.~6.

\section{Model and parameter grid}

In this section we describe the model used throughout the paper. The parameters for stars with different spectral types and activity levels are then reviewed.

\subsection{Simulation of magnetic structures and raw time series}

The approach is described in detail in \cite{borgniet15} for the Sun and in Paper I for other stars. The solar chromospheric model proposed in \cite{meunier18a} is also extended to other stars (Paper I).  This model is based on a large number of parameters (see next section) to produce spots, plages, and network structures in a consistent way. In this series of papers, we cover spectral types from F6 to K4 and relatively old main-sequence stars (maximum average LogR'$_{\rm HK}$ depending on spectral type, from -4.6 for F6 stars to -4.85 for K4 stars).

We recall here the time series used in the following analysis:
\begin{itemize}
\item{ {\it rvspot1} and {\it rvspot2:} RV due to the presence of spots, following two laws (see Sect.~2.2) for the spot temperature contrast (a lower and an upper limit, $\Delta$Tspot$_1$ and $\Delta$Tspot$_2$, respectively).}
\item{ {\it rvplage:} RV due to plages and network.   }
\item{ {\it rvconv:} RV due to the inhibition of the convective blueshift in plages and in the network. In the following, "convection" refers to this process, which is related to both granulation properties and to magnetic activity. This is different from the granulation signal {\it rvogs} described below.   }
\item{ {\it rvact1} and {\it rvact2}: sum of {\it rvspot1} ({\it rvspot2}), {\it rvplage,} and {\it rvconv}. }
\item{ {\it rvogs:} RV due to oscillation, granulation, and supergranulation (hereafter the OGS signal), either at the original temporal resolution (30 s) or  averaged over one hour. }
\item{ {\it rvinst:} RV due to instrumental white noise represented by a Gaussian noise with an amplitude of 0.6 m/s.}
\item{ {\it LogR'$_{\rm HK}$:} chromospheric emission index derived from the S-index produced in the simulation \cite[][]{harvey99,meunier18a}.}
\end{itemize}

The temporal step is on average one day (with random departures up to four hours). The time series cover an integer number of cycles, with a maximum of 15 years, the duration therefore varies between 3327 and 5378 days depending on the simulation. We also used a degraded sampling that is defined as follows:  a four-month gap each year was introduced to simulate the fact that a star is not always visible from a given observatory; a maximum duration of 9 years was imposed to degrade the duration, so that in some cases not a complete activity cycle was covered;  the  number of point $N_{\rm obs}$ was reduced to 100, 300, 500, 1000, or 2000 depending on the time series, with a random choice between these four values and a ramdom choice between 100 realizations of the sampling). 


\subsection{Parameter grid}


\begin{figure}[h!]
\includegraphics{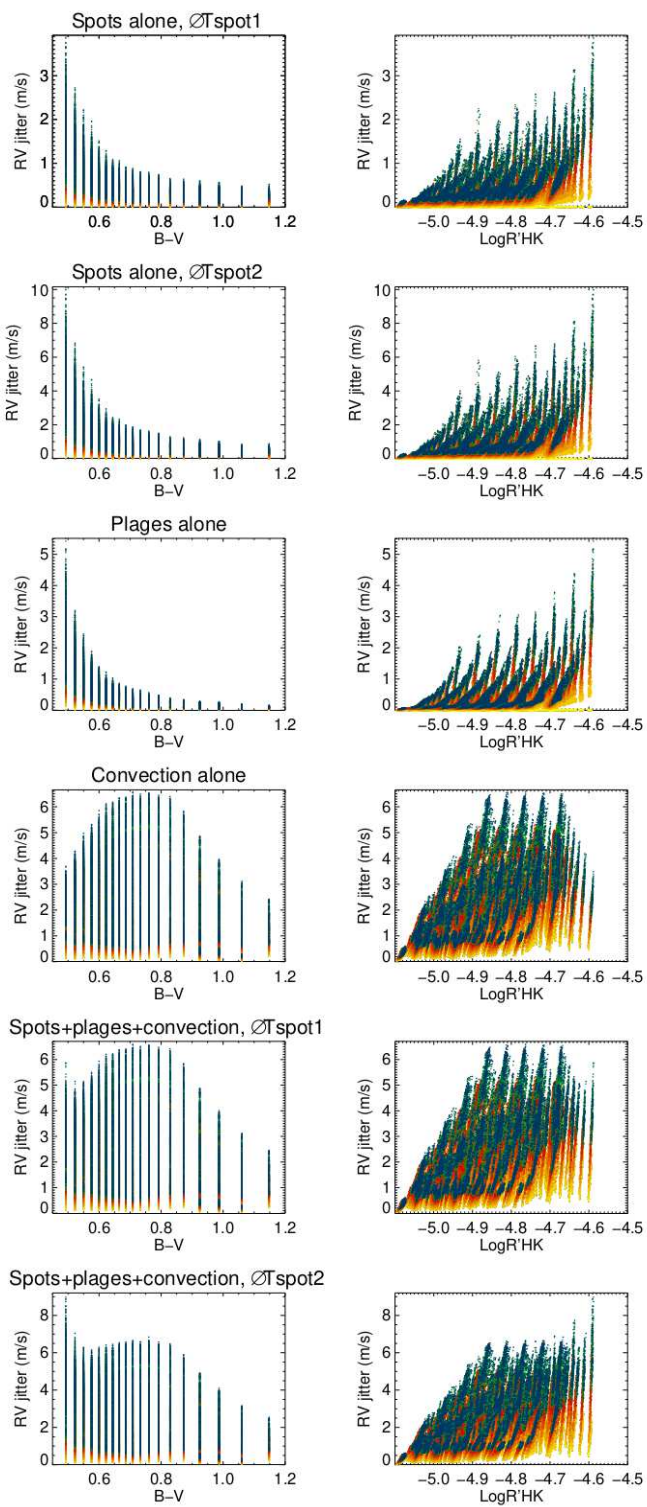}
\caption{
RV jitter for the different magnetic activity components to the RV, vs. B-V (left) and LogR'$_{\rm HK}$ (right), {\it from top to bottom}: {\it rvspot1}, {\it rvspot2}, {\it rvplage}, {\it rvconv}, {\it rvact1,} and {\it rvact2}. Each point corresponds to one simulation. The color code corresponds to the inclination, from pole-on (i=0$^{\circ}$, yellow) to edge-on (i=90$^{\circ}$, blue); here light and dark orange correspond to 20$^{\circ}$ and 30$^{\circ}$, light and dark red to 40$^{\circ}$ and 50$^{\circ}$, brown to  60$^{\circ}$, and light and dark green to 70$^{\circ}$ and 80$^{\circ}$. 
}
\label{composante}
\end{figure}


We list here the parameters that we modified from the solar values. More details about the laws, justification, and references can be found in Paper I. Our basic sets of parameters  depend on B-V and LogR'$_{\rm HK}$. For each point in this 2D  (B-V, LogR'$_{\rm HK}$) grid, the following parameters were adapted (from the solar values):
\begin{itemize}
\item{Fundamental parameters (T$_{\rm eff}$, mass, radius). They depend on the spectral type (and therefore B-V).}
\item{Spot temperature contrast. We used two laws: a lower limit defined by the solar contrast  in \cite{borgniet15}, $\Delta$Tspot$_1$, and an upper  limit law depending on T$_{\rm eff}$ from \cite{berd05}, $\Delta$Tspot$_2$.  We assumed that star spots have contrasts within this range. }
\item{The plage contrast. It depends on B-V, but also on the size and position of each structure \cite[][]{unruh99,meunier10a,borgniet15,norris18}. }
\item{Convective blueshift and attenuation factor. They depend on B-V and LogR'$_{\rm HK}$. We also scaled the diffusion coefficient to the convective blueshift amplitude \cite[][]{meunier17b}.  }
\item{The maximum average latitude at the beginning of the cycle $\theta_{\rm max}$. It is not constrained observationally or from models, therefore we studied the effect of three values: the solar latitude $\theta_{\rm max,\odot}$, $\theta_{\rm max,\odot}$+10$^{\circ}$, and $\theta_{\rm max,\odot}$+20$^{\circ}$. The two last values produce butterfly diagrams with a larger extent in latitude than in the Sun and were chosen arbitrarily in Paper I to study the effect of this important parameter.}
\item{The rotation period P$_{\rm rot}$. It depends on B-V and on the average LogR'$_{\rm HK}$. We considered three laws: a median law, a lower limit law, and an upper limit law, assuming that the observed dispersion in the relationship between P$_{\rm rot}$ and LogR'$_{\rm HK}$ reflects an actual dispersion between stars \cite[][]{mamajek08}. }
\item{The differential rotation. It depends on T$_{\rm eff}$, but also on the rotation period \cite[][]{reinhold15}, and on $\theta_{\rm max}$.  }
\item{The cycle period P$_{\rm cyc}$. It depends on the rotation period (and therefore also  on B-V and LogR'$_{\rm HK}$). We considered three laws (median law, and two extreme laws), assuming  that the observed dispersion in the relationship between the cycle period and P$_{\rm rot}$ is real \cite[][]{noyes84,baliunas95,saar99,bohm07,olah09,lovis11b,suarez16,olah16}.}
\item{The cycle amplitude A$_{\rm cyc}$. It depends on B-V and LogR'$_{\rm HK}$. We considered three laws: median law, and two extreme laws \cite[][]{lovis11b,radick98}. }
\item{OGS parameters. The typical frequencies and frequency width of the power spectra (for the oscillations), timescale, and exponent in the power spectra (granulation and supergranulation) as well as their amplitude depend on B-V \cite[][]{kallinger14,kjeldsen95,samadi07,bedding03,kippenhahn90,belkacem13,harvey84,beeck13,meunier15} .}
\end{itemize}

All other parameters were kept to our solar values  as in \cite{borgniet15}: meridional circulation \cite[][]{komm93}, other plage and network properties  \cite[][]{borgniet15,schrijver01,meunier17b}, and spot properties \cite[][]{martinez93,baumann05,borgniet15}.
The five parameters for which we tested several laws (spot contrast, $\theta_{\rm max}$, P$_{\rm rot}$, P$_{\rm cyc}$ , and A$_{\rm cyc}$)  were used to produce time series that correspond to the same (B-V, LogR'$_{\rm HK}$) point. We study the effect of these parameters in Sect.~3.2.
Finally, the inclination, although not an intrinsic parameter, strongly affects the RV jitter.

\section{Effect of parameters on the RV jitter}

\begin{figure}
\includegraphics{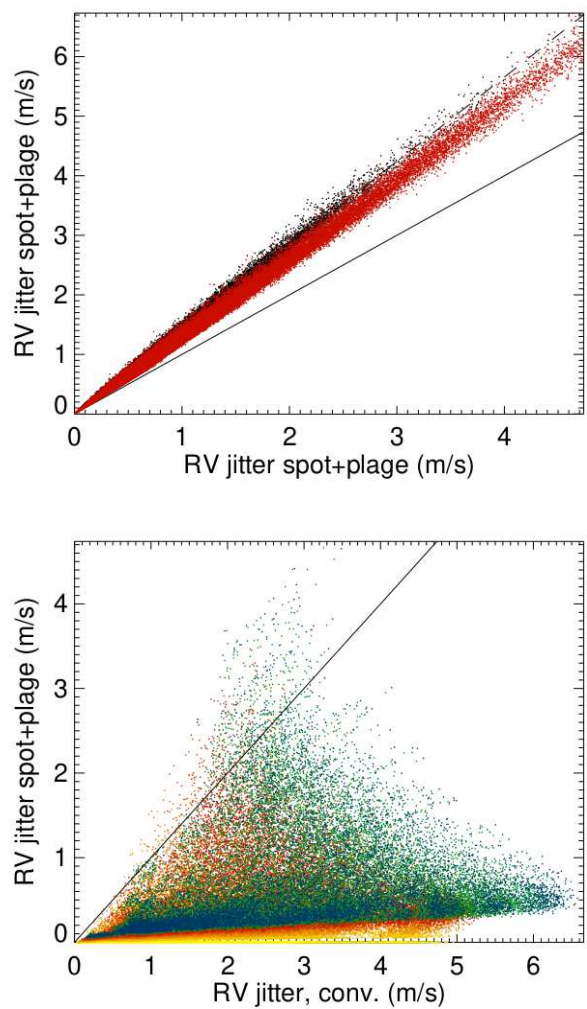}
\caption{
{\it Upper panel:}  Root mean square of the RV jitter due to spot ({\it rvspot1} in black and {\it rvspot2} in red)) and plages separately vs. the RV jitter of {\it rvspot}+{\it rvplage}. The solid line indicates a y=x linear function and the dashed line has a slope of $\sqrt{2}$. 
{\it Lower panel:} RV jitter of {\it rvspot1}+{\it rvplage} vs. the RV jitter of {\it rvconv}. The color code is similar to Fig.~1.  The RV jitter of simulations above the solid line arises because {\it rvspot1}+{\it rvplage} is larger than the RV jitter due to {\it rvconv}.
}
\label{composante2}
\end{figure}

\begin{figure}
\includegraphics{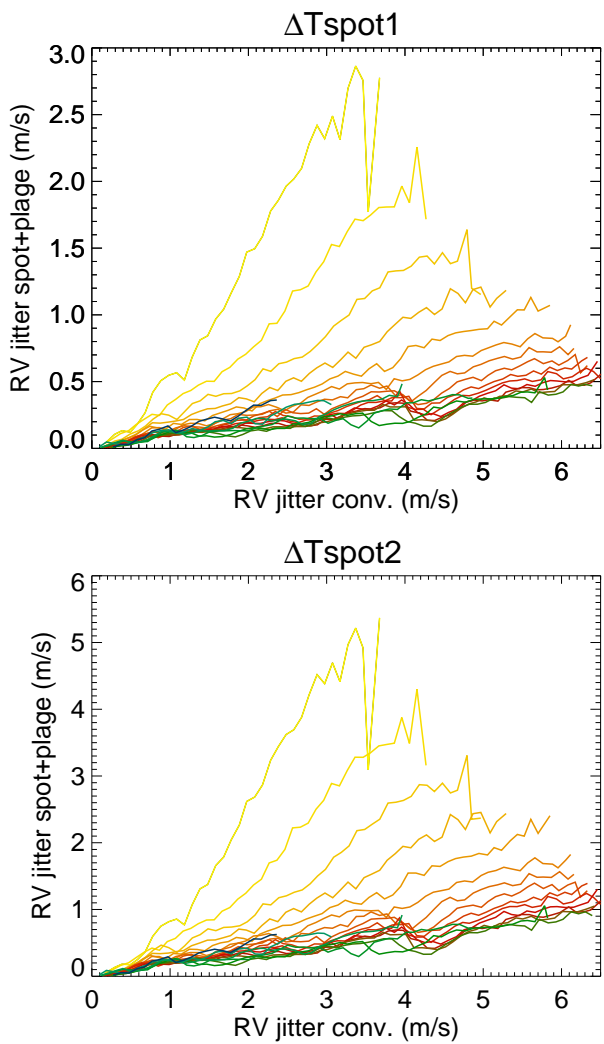}
\caption{
{\it Upper panel:} Binned RV jitter of {\it rvspot}+{\it rvplage} vs. RV jitter from the attenuation of the convective blueshift {\it rvconv} for the 19 spectral types from F6 (yellow) to K4 (blue). 
{\it Lower panel:} Same for $\Delta$Tspot$_2$.
}
\label{compbin}
\end{figure}

In this section, we study the effect of the parameters on the RV jitter to identify  the most critical parameters. This is important because many parameters have complex dependencies, some of which produce strong geometrical effects. 

\subsection{Magnetic contributions to the RV jitter}

We illustrate how the different components of magnetic activity (spots, plages, and inhibition of the convective blueshift in plages) listed in Sect.~2.1 depend on the stellar spectral type and LogR'$_{\rm HK}$ and how they relate to each other. We focus on the magnetic activity contributions without noise or OGS signal. Fig.~\ref{composante} shows a strong decrease in the RV jitter that is induced by spots and plages toward larger B-V and toward quiet stars. The main reason for the former result is likely to be the decrease in contrast (both for spot and plages). 
As expected, the RV jitter is larger for {\it rvspot2} compared to {\it rvspot1}. 
The convective component first increases with B-V and then decreases again. There is an expected decrease in amplitude of the convective blueshift itself, but this is modulated by the amplitude of the cycle, which is dominating the trend. 
The RV jitter on the complete time series is slightly higher  for the highest spot temperature contrast, but the difference between the final RV jitter for {\it rvact1} and {\it rvact2} is very small because {\it rvconv} often dominates. 
Finally, on average, the RV jitter (due to convection and total) is higher when seen edge-on compared to pole-on. This is mostly due to the rotational modulation effect and partly to the long-term variability (one-third). The ratio between the edge-on and pole-on RV jitter for  {\it rvconv} strongly varies; it takes values between 0.1 and 1.1 for the most extreme cases. 



In the following we illustrate our results with $\Delta$Tspot$_2$ unless indicated otherwise. 
We know that the signals from spots and plages partially compensate for each other because they are located at roughly the same positions \cite[as shown in][]{meunier10a}. Spots have a much higher contrast than plages, but are also much smaller. The compensation is not strict, however, because there is a dispersion in plage-to-spot size ratio and because plages have a longer lifetime. To illustrate this, Fig.~\ref{composante2} shows how the root mean square of the RV jitter due to spots and plages separately varies with the RV jitter of the {\it rvspot}+{\it rvplage} time series. They both lie between the two straight lines (the lowest line corresponds to an exact compensation between plages and spots), indicating some compensation, although they do no compensate entirely. The compensation is slightly better with $\Delta$Tspot$_2$. 

The lower panel of Fig.~\ref{composante2} shows how the RV jitter due to spots and plages varies with the RV jitter due to {\it rvconv}. Although higher {\it rvconv} variability tends to be associated with high {\it rvspot}+{\it rvplage} variability, there is a high dispersion: for medium {\it rvconv} jitter, we observe the highest  {\it rvspot}+{\it rvplage} jitter, for example. There is also a weak inclination effect. The solid line indicates the limit between RV jitter that is dominated by {\it rvspot}+{\it rvplage} (above the line) and by {\it rvconv} (below the line). For $\Delta$Tspot$_1$, they correspond to less than 1\% of the simulations, all for F6 stars. For $\Delta$Tspot$_2$, the RV jitter in almost 5\% of the simulations (all corresponding to F stars) is higher for {\it rvspot}+{\it rvplage} than for {\it rvconv}. This is illustrated in Fig.~\ref{compbin}, which shows a version of the lower panel of  Fig.~\ref{composante2} separately for each spectral type and after binning  in {\it rvconv}.

\subsection{Effect of cycle parameters (A$_{\rm cyc}$, P$_{\rm cyc}$, $\theta_{\rm max}$) and P$_{\rm rot}$ on RV jitter}

\begin{figure}
\includegraphics{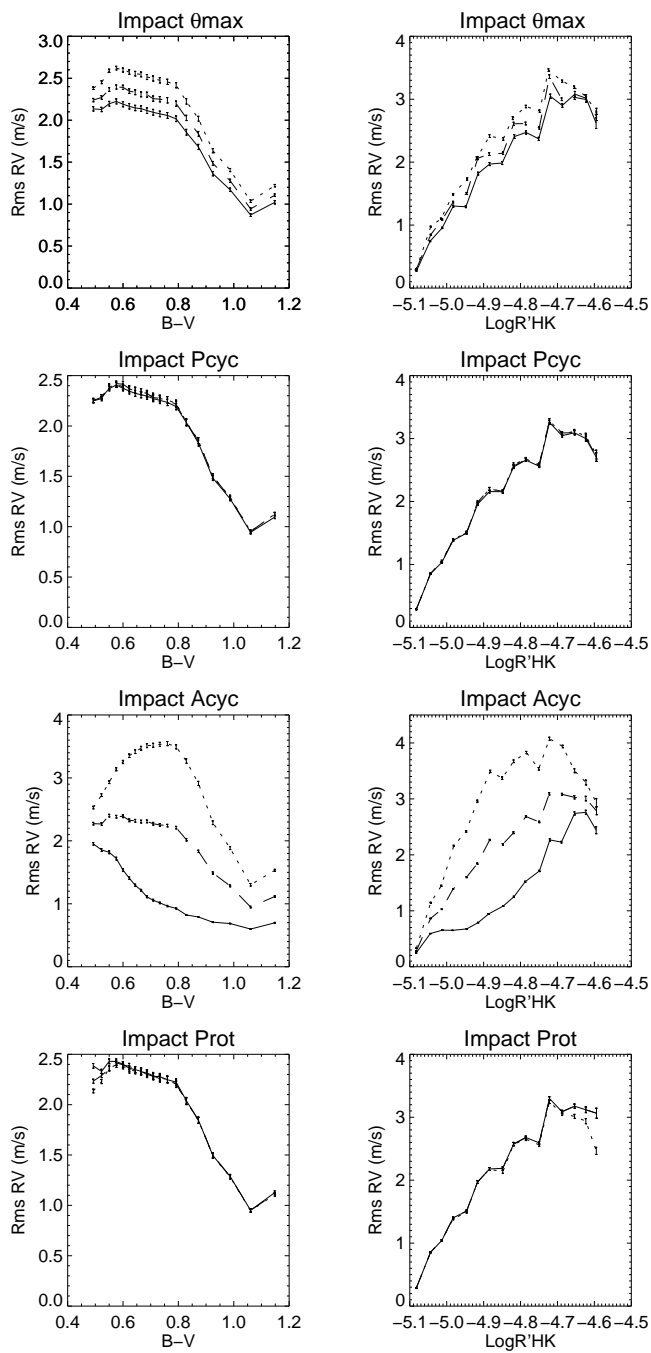}
\caption{
{\it First line:} Average RV jitter (computed on {\it rvact2} with no noise) in bins of B-V (left) and LogR'$_{\rm HK}$ (right), for the three levels considered for $\theta_{\rm max}$: lower value (solid), medium value (dashed line), and higher value (dotted line).
{\it Second line:} Same for the effect of P$_{\rm cyc}$.
{\it Third line:} Same for the effect of A$_{\rm cyc}$.
{\it Fourth line:} Same for the effect of P$_{\rm rot}$.
}
\label{param}
\end{figure}

\begin{figure}
\includegraphics{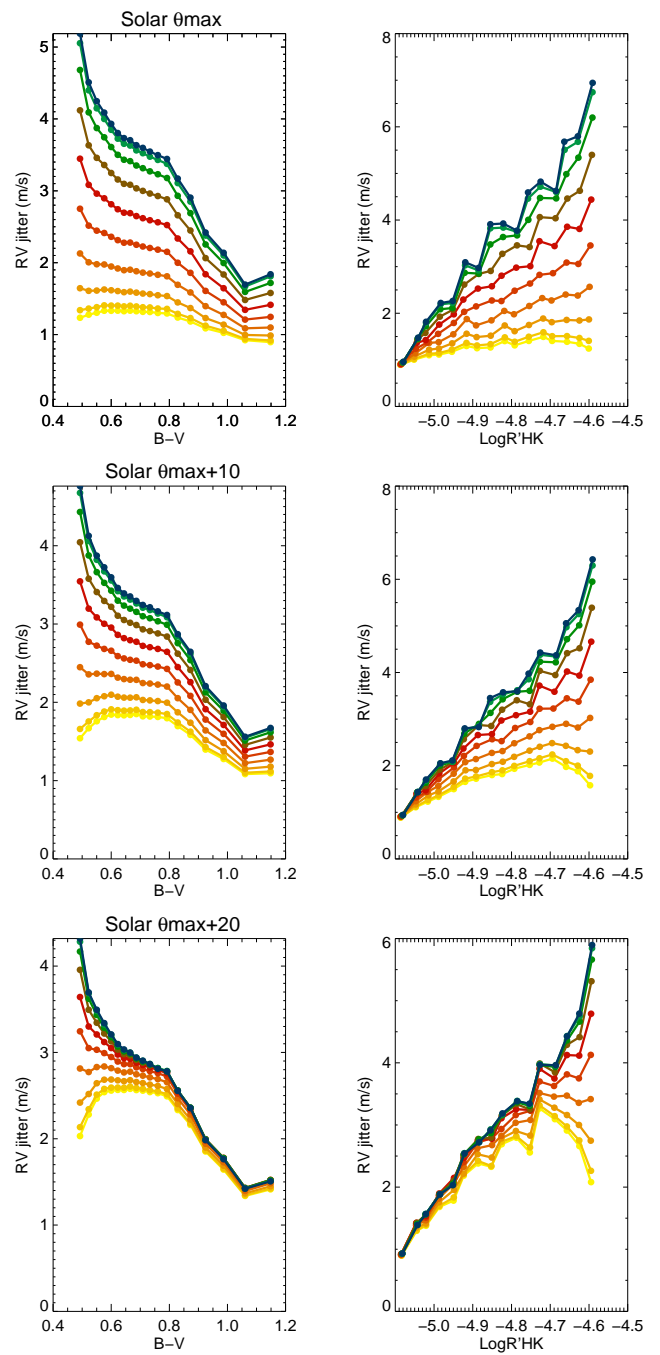}
\caption{
{\it Upper panel:} RV jitter (computed on {\it rvact2} added to {\it rvogs} and {\it rvinst}) vs. B-V (right) and LogR'$_{\rm HK}$ (left), for $\theta_{\rm max}$ equal to the solar value, and after binning in B-V and LogR'$_{\rm HK}$ , respectively. The color code is similar to Fig.~1. 
{\it Middle panel:} Same for $\theta_{\rm max}$ equal to the solar value +10$^{\circ}$.
{\it Lower panel:} Same for $\theta_{\rm max}$ equal to the solar value +20$^{\circ}$.
}
\label{latmax}
\end{figure}

\begin{figure}
\includegraphics{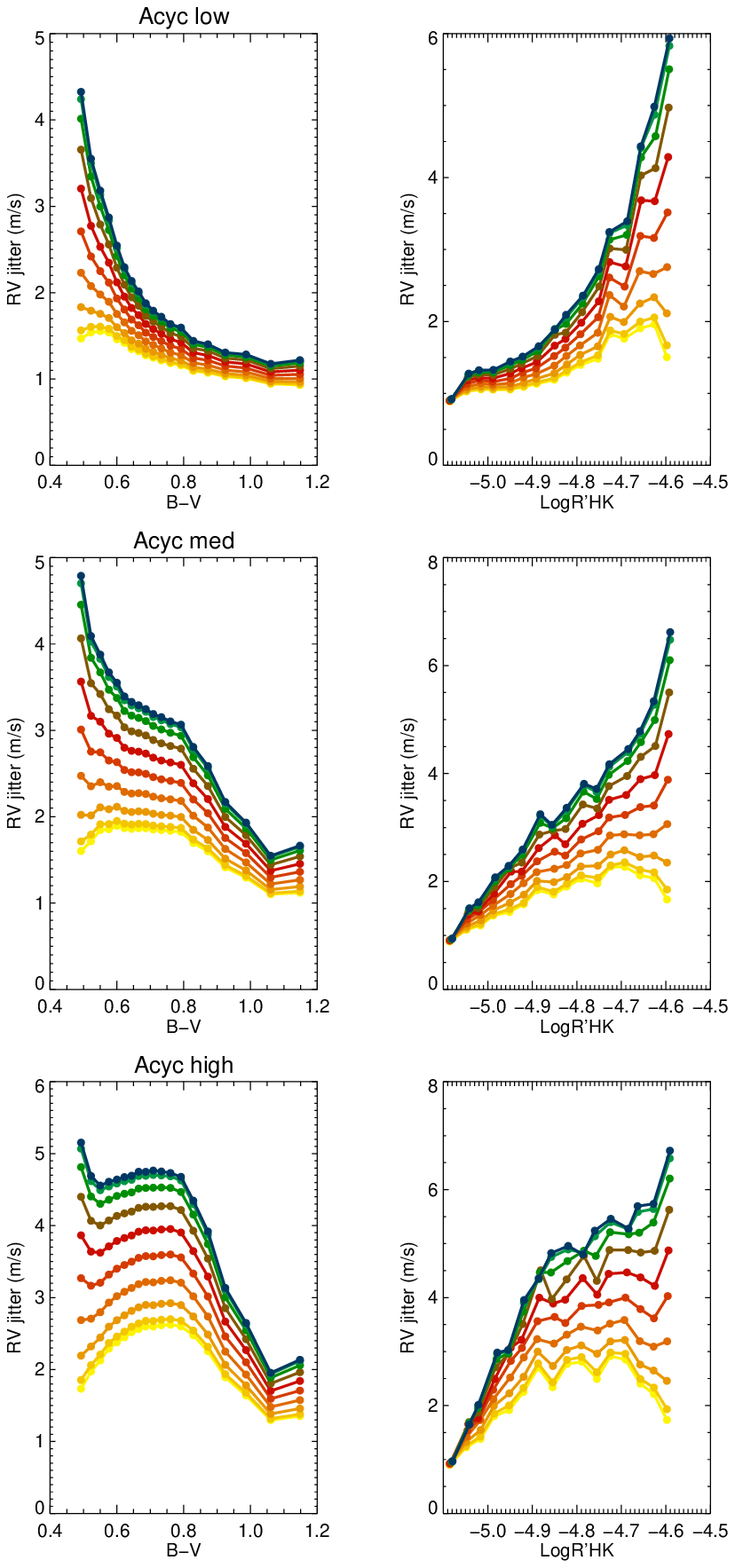}
\caption{
Same as Fig.~5 for A$_{\rm cyc}$.
}
\label{acyc}
\end{figure}

We now consider the time series including all contributions (magnetic, OGS averaged over one hour, and instrumental) for $\Delta$Tspot$_2$. The results are very similar for $\Delta$Tspot$_1$ and when {\it rvact1} or {\it rvact2} are considered alone (not shown here).

In the previous section, we have shown the dependence of the different components of the RV jitter on B-V and LogR'$_{\rm HK}$ and discussed the effect of $\Delta$Tspot. We now consider the four other laws for which we test different levels (see Sect.~2.2): $\theta_{\rm max}$, P$_{\rm cyc}$, A$_{\rm cyc}$, and P$_{\rm rot}$. 
Fig.~\ref{param} shows a global view of the effect of these parameters after binning in B-V and LogR'$_{\rm HK}$, providing one curve for each level. We note that the different choices of $\theta_{\rm max}$ do not depend on the other parameters (they take either the solar value or a value of +10$^{\circ}$ and +20$^{\circ}$), while for the other three parameters, all laws (lower, medium, and higher) vary with the other parameters (mainly spectral type and activity level). P$_{\rm rot}$ and P$_{\rm cyc}$  do not significantly affect the RV jitter, at least on average: this is not surprising because we expect them to  strongly affect the frequency dependence of the variability, but not its global amplitude (the frequency analysis will be the subject of a future paper). On the other hand, $\theta_{\rm max}$ and A$_{\rm cyc}$ strongly affect  the RV jitter, the first because of geometrical effects (e.g., inclination), and the second naturally through the number of structures that are injected. 


We therefore study the effect of $\theta_{\rm max}$ and A$_{\rm cyc}$ in more detail in Figs.~\ref{latmax} and~\ref{acyc}, respectively. 
These plots show the RV jitter separately for the three laws versus B-V and LogR'$_{\rm HK}$. They simultaneously show the inclination effect. 
We observe that the weak increase in RV jitter when $\theta_{\rm max}$ increases masks a larger diversity that is due to inclination because the interplay between $\theta_{\rm max}$ and inclination is strong. For the lower $\theta_{\rm max}$ values (upper panels), the RV jitter strongly increases with inclination. However, for higher $\theta_{\rm max}$ values, the effect of inclination is much weaker and remains strong only for F stars. Fig.~\ref{acyc} shows similar plots for A$_{\rm cyc}$. On average, we observe higher RV jitter at high inclinations at all B-V, mostly for high-mass stars. The inclination effect is stronger for high cycle amplitudes.




\section{RV jitter versus B-V and LogR'$_{\rm HK}$}

In this section, we compare RV with observations, first for the RV jitter, and then for the slope between RV and R'$_{\rm HK}$. Finally, we focus on the comparison between short-term and long-term correlations between RV and LogR'$_{\rm HK}$.

\subsection{Comparison of the RV jitter with observations}

\begin{figure}[h!]
\includegraphics{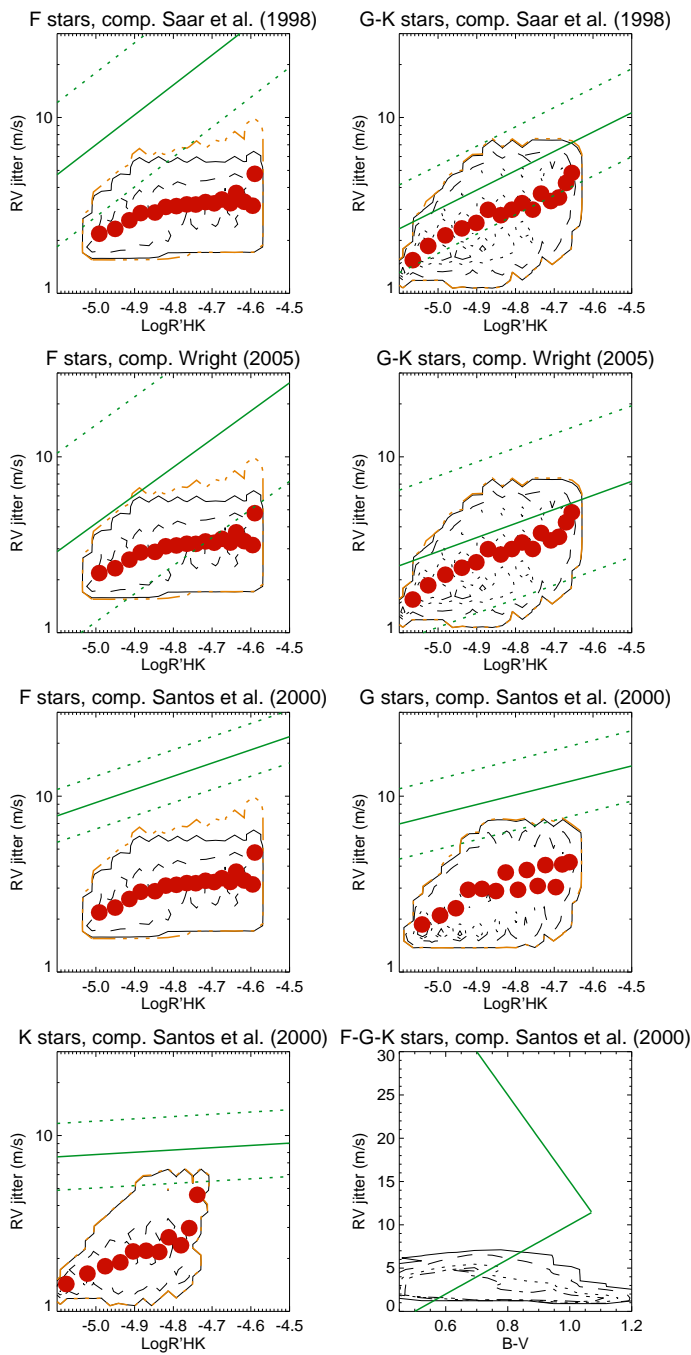}
\caption{
{\it First line:} RV jitter for F stars (left) and G+K stars (right)  vs. LogR'$_{\rm HK}$ for all our simulations, in red (averaged in LogR'$_{\rm HK}$ bins, $\Delta$Tspot1), black contours (levels for 1, 100, 500, and 1000 simulations, $\Delta$Tspot1), and orange contours (level-one simulation, $\Delta$Tspot2). The green solid line indicates the law derived by Saar et al. (1998), and the two dashed green lines indicate the approximate range of RV jitter that was covered by their observations.  
{\it Second line:} Same with a comparison with Wright (2005).  
{\it Third and fourth lines:} Same with a comparison with Santos et al. (2000) for the first three panels (F, G, and K separately). The last panel represents the entire RV jitter vs. B-V. The two green lines indicate the lower and upper bounds of RV jitter values found by Santos et al. (2000). 
}
\label{rvjitter}
\end{figure}

In this section, we use the series we generated when the contribution from magnetic activity was taken into account, as well as the oscillations, granulation, and supergranulation signal (with no averaging to represent realistic short-term exposures). We do not include instrumental noise here because the observed RV jitter with which we compare our jitter is assumed to be corrected for them. Time exposures should be relatively short (and in any case much shorter than one hour), although they are not available. 
The RV jitter was computed on the whole time series, and therefore includes signal at all temporal scales. It is therefore not representative of what happens at a specific scale, but instead includes different contributions, which may vary in different proportions from one series to the other. 

Fig.~\ref{rvjitter} shows a comparison of our RV jitter with the jitter provided in three publications \cite[][]{saar98,santos00,wright05} that all studied a large sample of F-G-K stars.  
For G and K stars, the trend versus LogR'$_{\rm HK}$ is similar between observations and simulations, with a higher  RV jitter for high-activity level. The RV jitter also decreases toward lower mass stars in simulations and observations. The amplitudes are often smaller in our simulations, however, especially at low-activity level. We find that we cannot reproduce RV jitter  higher than 2-3 m/s with activity, oscillation, granulation, and supergranulation combined for very quiet stars. At high-activity level the agreement is good, especially with the RV obtained by  \cite{wright05}. 

However, it is difficult to understand how a  higher variability in RV  due to these effects (spots, plages, inhibition of the convective blueshift, and OGS) could be significantly  higher for such stars without increasing their average LogR'$_{\rm HK}$. 
Our interpretation is that the discrepancy between simulations (lack of very quiet stars with RV jitter in the 5-10 m/s range) and observations (significantly  higher RV jitter) could have three possible causes: 
\begin{itemize}
\item{{\it \textup{The instrumental contribution has been underestimated in these publications}.} This is probably at least partially the case because depending on the reference, the RV jitter level is different. For example, \cite{wright05} found a much lower RV jitter than the other references: it was lower by a factor 2-3. RV jitter has also been produced by \cite{isaacson10} for a large sample of stars that we do not show here. Their trend with activity level is less clear than for the three previous works, but the authors provide a lower bound that is also higher than our lower bound.  
}
\item{{\it \textup{The observed RV jitter is not only caused by activity and short-term stellar variability, but also by other sources}}, for example, still-undetected planets and binaries. This is consistent with the HARPS  data, which suggest that the RV jitter is overestimated due to activity in previous publications. 
We looked at a sample of stars observed with HARPS and with a low-activity level (LogR'$_{\rm HK}$ lower than -5).
Out of a sample of 408 F-G-K stars with LogR'HK lower than -5.54 (to fit our range of parameters), 35 stars have an average LogR'$_{\rm HK}$ lower than -5.0 and an RV jitter (including instrumental noise) larger than 4 m/s (i.e., they lie outside our simulation results). A close examination of these 35 stars shows that this RV jitter is mostly due to binaries (21 stars) and to published planet(s) (9 stars). We examined the six remaining stars in detail. Four stars show long-term variations that might be due to one (or more) planet, although the PI of the observations did not published them yet (the RV signal at long period for three stars is not correlated with LogR'$_{\rm HK}$ and is significantly above the false-alarm probability, while for the fourth the RV signal is not yet significant). For another star the dispersion in RV appears to occur mostly at very short timescales (minutes), for example, with a jump of 5 m/s over a few minutes, which is beyond the scope of our simulation and for which we have no explanation: the star is a G2 star and is expected to exhibit much smaller oscillations. The last star presents variations at the scale of a few dozen days for both RV and LogR'$_{\rm HK}$, with indeed relatively large RV variations, but there are also jumps at very short scales. In conclusion, HARPS data, which have a much better uncertainties than those of previous instruments, show that very few stars are expected to populate the domain with low-activity level and RV jitter higher than a few m/s, in agreement with our simulations, except for companions or instrumental noise. 
}
\item{{\it \textup{The parameters of the simulations are not well adapted to the samples analyzed in these papers}.} Mostly three parameters might cause a significant increase in the RV jitter: the number of structures, the size of the structures, and their contrast (for the {\it rvspot1}, {\it rvspot2}, and {\it rvplage} components) or amplitude of the inhibition of the convective blueshift (for the {\it rvconv} component). It is extremely unlikely that the number of structures should be underestimated by such a large factor because the Sun, seen edge-on (i.e., with maximum variability), is well modeled, and although its  LogR'$_{\rm HK}$ is well above -5, it does not have an rms larger than 3 m/s \cite[][]{meunier10a,meunier10}: The number of structures that would be required to reach more than a few m/s would then be incompatible with a very low LogR'$_{\rm HK}$ value. Similarly, the size of the structures, although not well constrained, is unlikely to be much larger than considered in our model because there is a trend to have larger spots for very active stars and conversely smaller spots for less active stars. Larger plages would lead to a larger LogR'$_{\rm HK}$. Finally, the inhibition of the convective blueshift has been well determined in \cite{meunier17b} and again is in good agreement for the Sun, which is more active than the quiet stars we discussed here (LogR'$_{\rm HK}$ below -5). In addition, it would not be compatible with the slope we observe between RV and LogR'$_{\rm HK}$ either, which agrees well with observations (see below).}
\end{itemize}

For F stars, the trend versus LogR'$_{\rm HK}$ we obtain is much flatter, and the discrepancy between the simulated and observed jitter is larger than for G and K stars. The reason might be that  our computations did not take the possible trend in convective blueshift attenuation with T$_{\rm eff}$ discussed in Sect.~3.2.1 into account. If this were included, the RV jitter would be slightly higher, although it would remain below the average found by \cite{wright05}. This shows that our simulations are quite robust with respect to the choice of parameters. As for the trend, the relatively flat curve is related to the fact that for F stars, cycle amplitudes are lower than for less massive stars, as discussed in Sect.~2.6.2: even for high-activity levels, the cycle amplitudes are therefore not as different from the lower activity level amplitudes. Therefore a much higher RV jitter at high-activity level is not compatible with such low cycle amplitudes. One possible explanation could be that the magnetic flux in quiet regions could depend differently on the activity level for G or K stars, which is unknown.

\subsection{Comparison of the RV versus R'$_{\rm HK}$ slope with observations}

\begin{figure}
\includegraphics{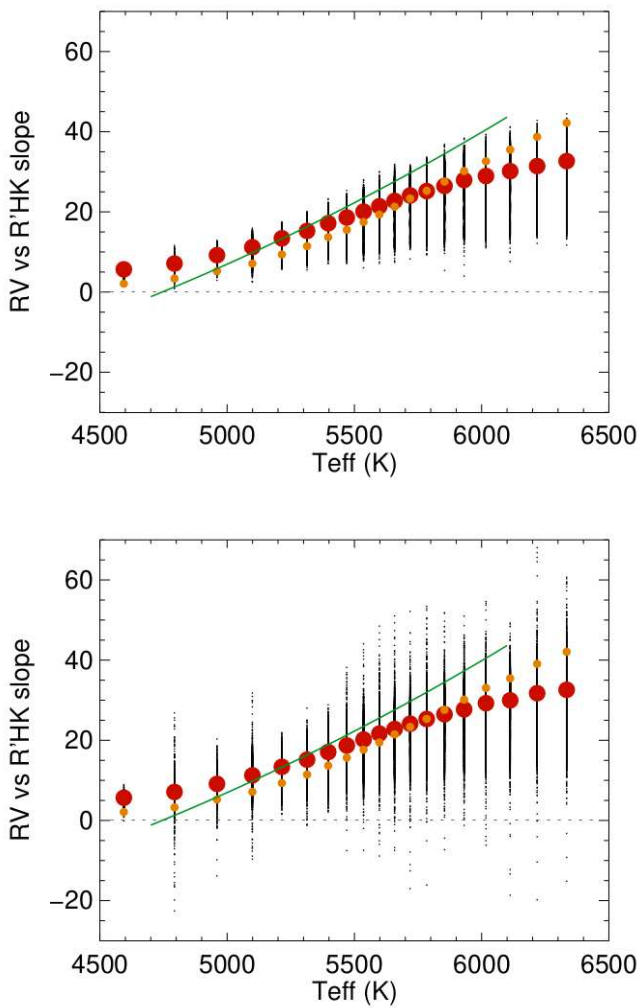}
\caption{
{\it Upper panel:} Slope of RV vs. R'$_{\rm HK}$, shown as a function of T$_{\rm eff}$. The red points indicate the average slope for each T$_{\rm eff}$ bin. The green line is the trend fit to observations in Lovis et al. (2011). The orange points correspond to the average when the trend vs. T$_{\rm eff}$ in the convective blueshift inhibition law (which we estimate is not valid below 5300~K) is taken into account.  
{\it Lower panel:} Same with degraded sampling, either 100, 500, 1000, or 2000 points depending on the simulation (see text). 
}\label{slope_rv_ca}
\end{figure}

We now consider time series that are constituted by magnetic activity, OGS (averaged over one hour or not) and instrumental noise, with the original sampling and a degraded sampling. 
We compute the slope of RV versus R'$_{\rm HK}$ to compare with the results of \cite{lovis11b}. This slope  can then be visualized as a function of other parameters, for example, T$_{\rm eff}$. Because the inhibition of the convective blueshift plays an important role and because both the chromospheric emission and this contribution to the RV are strongly related to the surface covered by plages (and network), we expect this slope to be significant and representative of the variability of these two variables. 

The slope is shown in Fig.~\ref{slope_rv_ca} versus T$_{\rm eff}$. This slope was also computed on a large number of F-G-K stars observed with HARPS by \cite{lovis11b}, and we compare their trend with our results. They show a good agreement, both in average trend (green line) and dispersion (although the dispersion is smaller in the simulations). This means that both our RV model (as far as the inhibition of convection contribution is concerned) and the chromospheric emission model (for the part due to active regions) are consistent with each other. There is a small departure for stars with either a very high mass or a very low mass (in our range of parameters), however. There is a possibility for the inhibition of the convective blueshift to depend slightly on T$_{\rm eff}$ \cite[][]{meunier17b}, at least above 5400~K, but it does not explain the discrepancy, although the slope is increased.  This comparison suggests that the trend (of the percentage of inhibition as a function of T$_{\rm eff}$) could be higher than expected from the results of \cite{meunier17b}, but this is not the only possible explanation. For example, in Paper I, we considered a similar chromospheric emission for all spectral types (for a given structure size) because this is not constrained. A large chromospheric emission (for a given size) for F stars and a lower chromospheric emission for K stars (compared to the Sun) could also increase the slope to better match the results of \cite{lovis11b}.  

Furthermore, even if all slopes are positive when the sampling is perfect, we occasionally observe negative slopes when the sampling is degraded. A few stars with a negative slope have been observed \cite[][]{lovis11b}. Lovis and collaborators have argued that these stars could present a reverse convection pattern (leading to a convective redshift instead of a convective blueshift), but this is unlikely to be the explanation because the convective properties of this same sample of stars have also been studied by \cite{meunier17b}, and none showed any reversal.  No such analysis was done by \cite{lovis11b}, whose objective was to concentrate on the variability derived from HARPS data, but \cite{meunier17b} derived the amplitude of the convective blueshift (including its sign). Their Fig.~9 also shows a few stars with a negative $\Delta$RV for a positive $\Delta$LogR'$_{\rm HK}$  \cite[similar to the results of][]{lovis11b}, but their Fig.~3 shows that the convective blueshift of all these stars has the same sign, which is not compatible with a convection reversal.
We propose that the few negative slopes that are observed might be an artifact due to poor sampling, although we cannot exclude that they are due to the presence of still-unidentified exoplanets.

\subsection{Short-term and long-term contributions to RV jitter}

\begin{figure*}[h]
\includegraphics{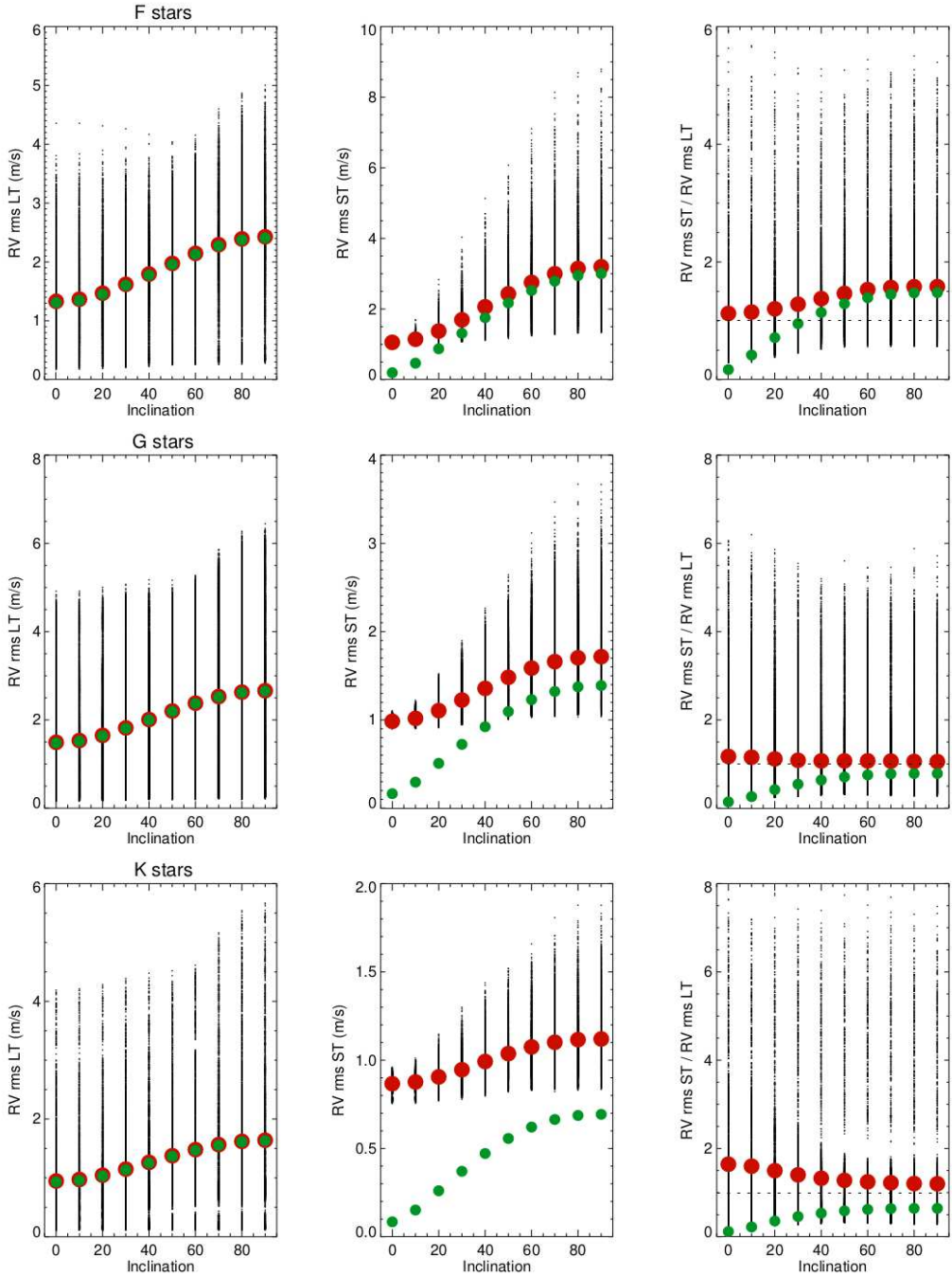}
\caption{
{\it Left column:} Long-term RV jitter vs. inclination (all simulations, black dots, and average, red points) for F stars (upper panel), G stars (middle panel), and K stars (lower panel) for activity and OGS contributions. The green points are the average values for activity alone. 
{\it Middle column}: Same for the short-term RV jitter. 
{\it Right column}: Same for the ratio between short-term RV jitter and long-term RV jitter. 
}
\label{ctlt}
\end{figure*}

\begin{figure*}[h]
\includegraphics{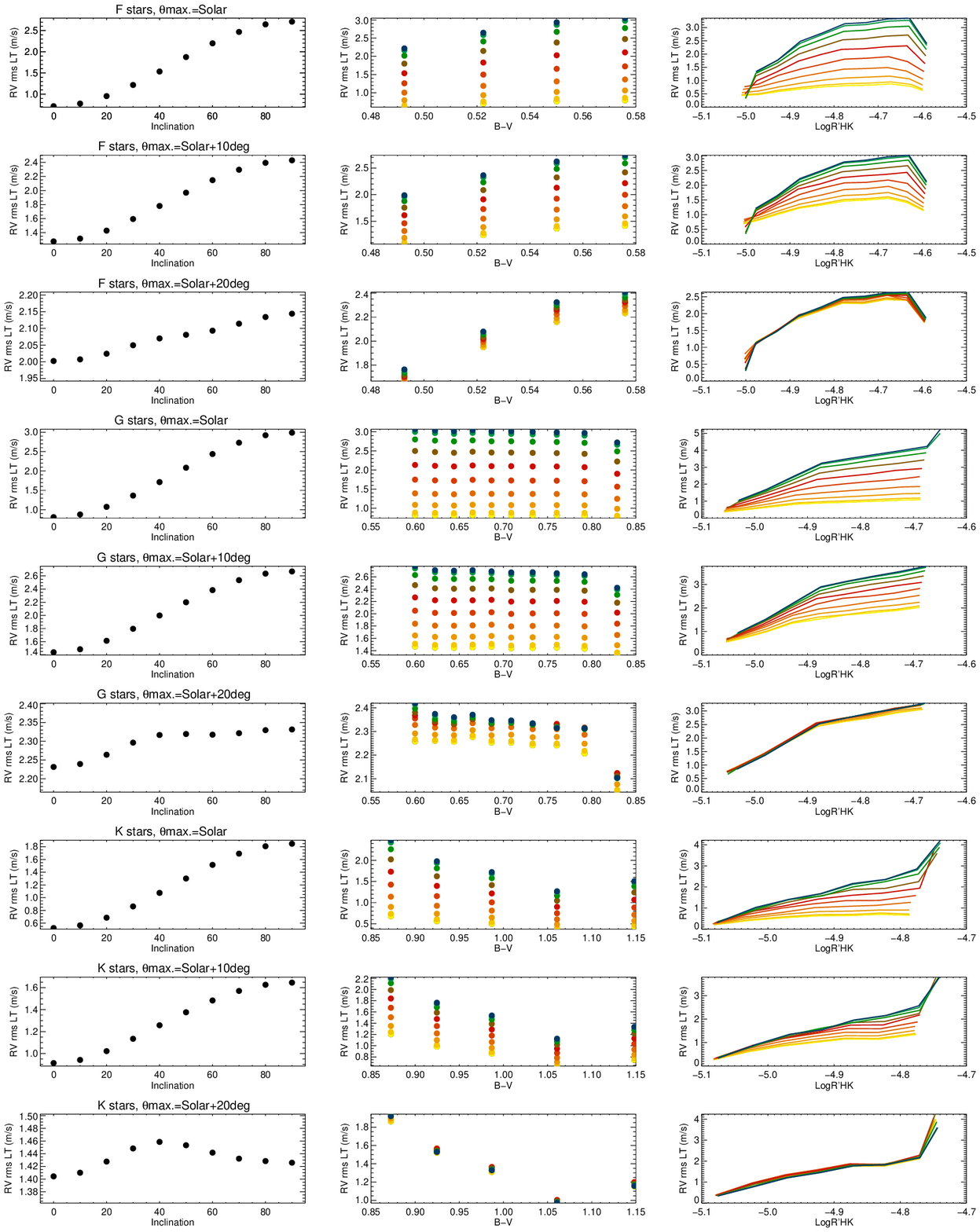}
\caption{
{\it Rows 1-3:} Binned long-term RV jitter vs. inclination (left) vs. B-V for various inclinations (middle) and vs. LogR'$_{\rm HK}$ for various inclinations (right) for F stars. The color code is similar to Fig.~1. The different rows correspond to different $\theta_{\rm max}$. 
{\it Rows 4-6:} Same for G stars.
{\it Rows 7-9:} Same for K stars.
}
\label{lt_theta}
\end{figure*}

The components {\it rvspot1} (as well as {\it rvspot2}) and {\it rvplage} are relatively short-term signals, that is, around the rotational period and below (although the dispersion changes on long timescales), while {\it rvconv} presents variations both at these short timescales and on long timescale (cycle variations). The RV jitter includes all components, short and long term.  Observationally, RV jitter also usually includes all components, hence the present study.  
We computed two cases of RV jitter, one on long timescales, hereafter LT (RV time series are smoothed over 100 days to eliminate most of the rotational modulation), and one on short timescales, hereafter ST (computed on the residuals after subtraction of the smoothed series).  The LT RV jitter is  mostly sensitive to {\it rvconv}, while the ST RV jitter is sensitive to all magnetic RV components ({\it rvspot}, {\it rvplage}, {\it rvconv}). 
We first illustrate the result with the series including the magnetic component (with $\Delta$Tspot$_2$), OGS (averaged over one hour), and instrumental white noise in Fig.~\ref{ctlt}. 
Long- and short-term RV jitter both increases with inclination. This is expected because the rotational modulation becomes increasingly weaker  from edge-on to pole-on. The long-term RV jitter is affected  by projection effects, and when seen edge-on, the structures present a larger apparent area. The ratio between short- and long-term contributions is on average always above 1 (i.e., dominated by short-term RV jitter), but the dispersion is very high and both configurations exist (i.e., dominated by either short term or long term). Furthermore, the ratio increases with inclination for F stars, is flat for G stars, and decreases with inclination for K stars. 

When we consider the magnetic component alone, the LT RV jitter is almost exactly the same (as expected), while the ST RV jitter is lower and is more strongly dependent on inclination because the OGS signal and instrumental noise do not depend on inclination.  As a consequence, the ratio between ST and LT RV jitter is quite different: it always increases with inclination, and is on average below one (except for F stars with inclinations higher than 40$^{\circ}$). The inclination dependence  therefore strongly depends on the other sources of signal (OGS and instrumental noise)  that are superposed on the activity contribution.


 We now focus on the long-term RV jitter, which is of interest for exoplanet detections in the habitable zone of the stars we consider. Fig.~\ref{lt_theta} shows the long-term jitter separately for the three values of $\theta_{\rm max}$. The jitter becomes flatter with inclination when $\theta_{\rm max}$ increases. For K stars and higher $\theta_{\rm max}$, there is even a reversal in the trend as the RV jitter passes through a maximum for inclination around 40$^{\circ}$. The effect of inclination becomes very small for the highest value of $\theta_{\rm max}$.  After binning, the RV jitter for F and G stars is not very different, and over all K stars is more suitable for exoplanet detection. The ratio between short- and long-term contributions depends on the spectral type: it is stronger for F stars when seen edge-on than for K stars. It is the opposite when seen pole-on.


We conclude that it is important to be careful in interpreting observed RV jitter because many configurations are present, depending on spectral type or inclination, for example. In addition, it is not possible to easily deduce a long-term variability from short-term observations, for example, and vice versa, because of the large diversity of configurations.

\section{Exoplanet detectability}

This section focuses on exoplanet detectability. We first describe the approach we have followed to derive mass detection limits. The criterion is then applied to derive typical planetary amplitudes and masses versus B-V and LogR'$_{\rm HK}$.

\subsection{Approach}

In this section, we use these simulations to predict exoplanet detectability in a simple way. Detailed and precise computations of detection limits, which request taking the frequency behavior of the time series into account, is beyond the scope of this paper and will be presented in a future study. To do so, we used the results of the fitting challenge performed by \cite{dumusque16} and \cite{dumusque17}. They proposed a criterion to establish a relation between the properties of a time series (jitter and number of observation namely) and the RV amplitude of a planet that can be detected assuming these properties. The criterion $C$ is defined as $C=K_{\rm pl} \times \sqrt{N_{\rm obs}} / \sigma$, where $K_{\rm pl}$ is the RV amplitude of the planet, $N_{\rm obs}$ is the number of observations, and $\sigma$ is the rms of the RV signal after a correction using a linear dependence on LogR'$_{\rm HK}$ and a second-degree polynomial in time (the latter is often performed on observed RV time series to remove any far and not well-constrained companion, either stellar or substellar). The fitting challenge \cite[][]{dumusque17} shows that the rate of success in recovering injected planets depended on $C$, with a very rough limit around 7.5: they showed that that current correction techniques were usually unsuccessful below this limit. We therefore considered $C$=7.5, computed the RV jitter $\sigma$ after the correction described above, and deduced the minimum $K_{\rm pl}$ for each time series, knowing the number of points.  From this value of $K_{\rm pl}$, a planet mass was  derived for a given period (we assumed eccentricities of 0). This provides an approximate detection limit across our grid of parameters that is representative of the current status of correcting techniques. 

We considered several time series in this computation, either the full sample (the number of points then vary depending on the simulation), or subsamples of 100, 300, 500, 1000, or 2000 points, as described in Sect.~2.1. The computations were made on the time series with OGS (averaged over one hour) and the instrumental white noise. 

\subsection{RV amplitude after correction}

\begin{figure}
\includegraphics{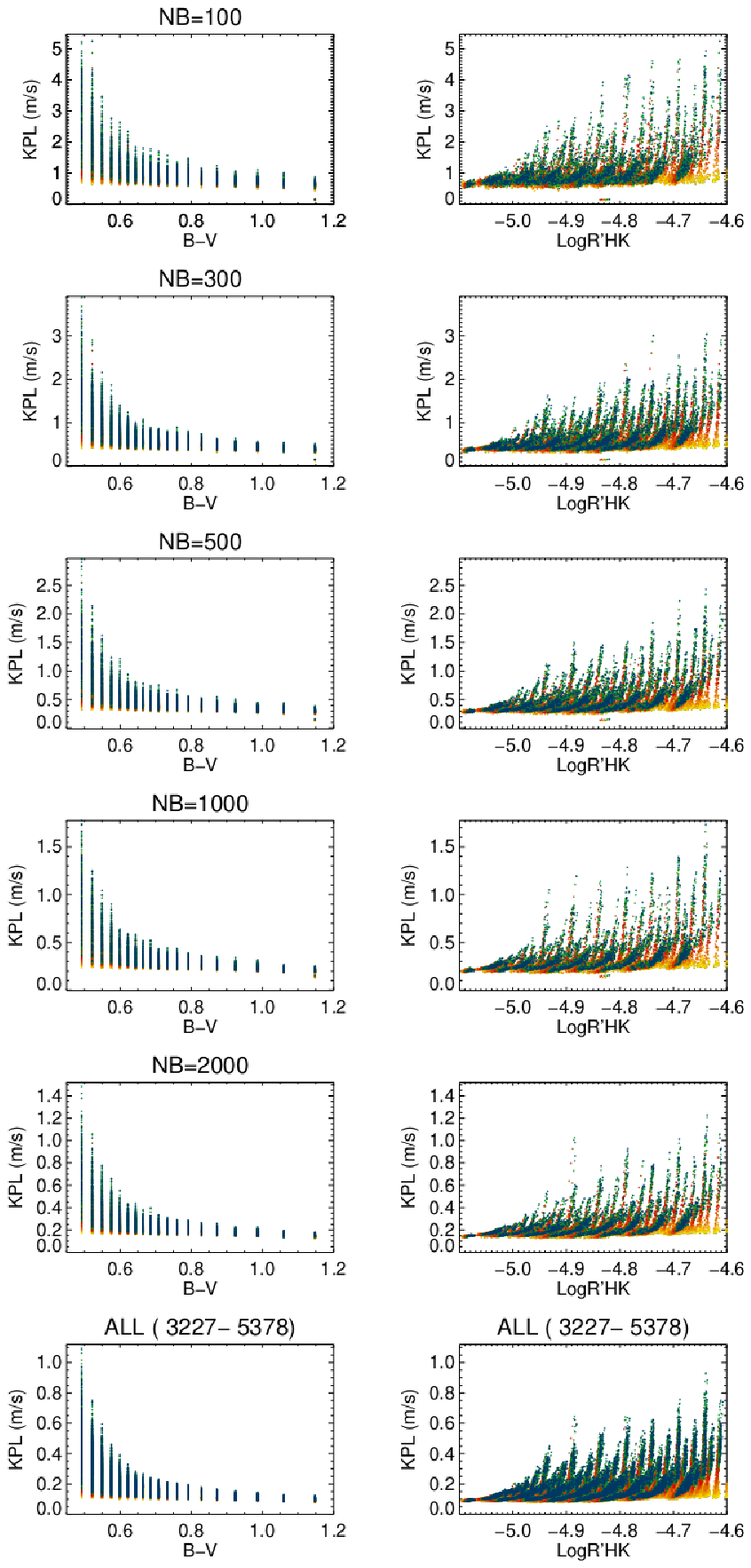}
\caption{
$K_{\rm pl}$ vs. B-V (left) and vs. LogR'$_{\rm HK}$ (right) for different numbers of points, from top to bottom: 100, 300, 500, 1000, 2000 and all points in the original time series. Computations are made on time series including {\it rvogs} and {\it rvinst} for $\Delta$Tspot$_2$ and $C$=7.5. Same color code as Fig.~1. 
}
\label{kpl}
\end{figure}

\begin{figure}
\includegraphics{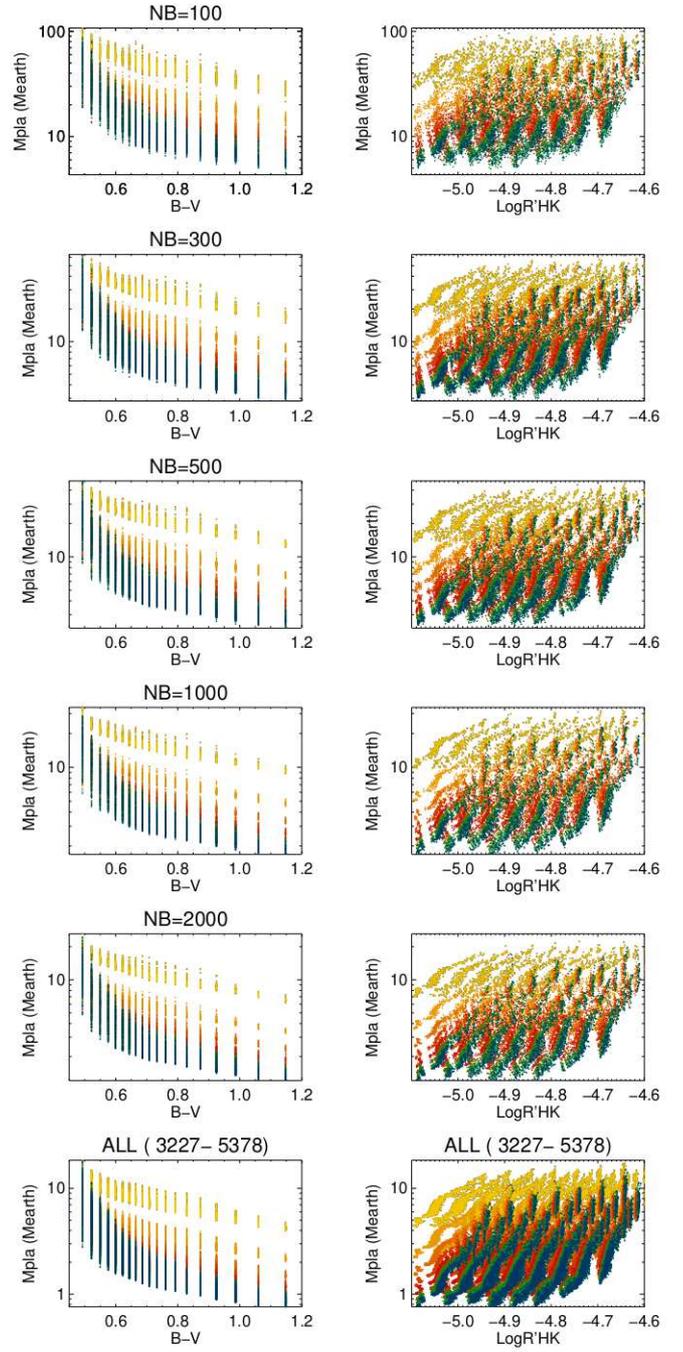}
\caption{
Same as Fig.~11 for M$_{\rm pla}$, after correction for sin(i) (simulations with i=0 are not included in this plot due to the correction). 
}
\label{mpla}
\end{figure}

\begin{figure}
\includegraphics{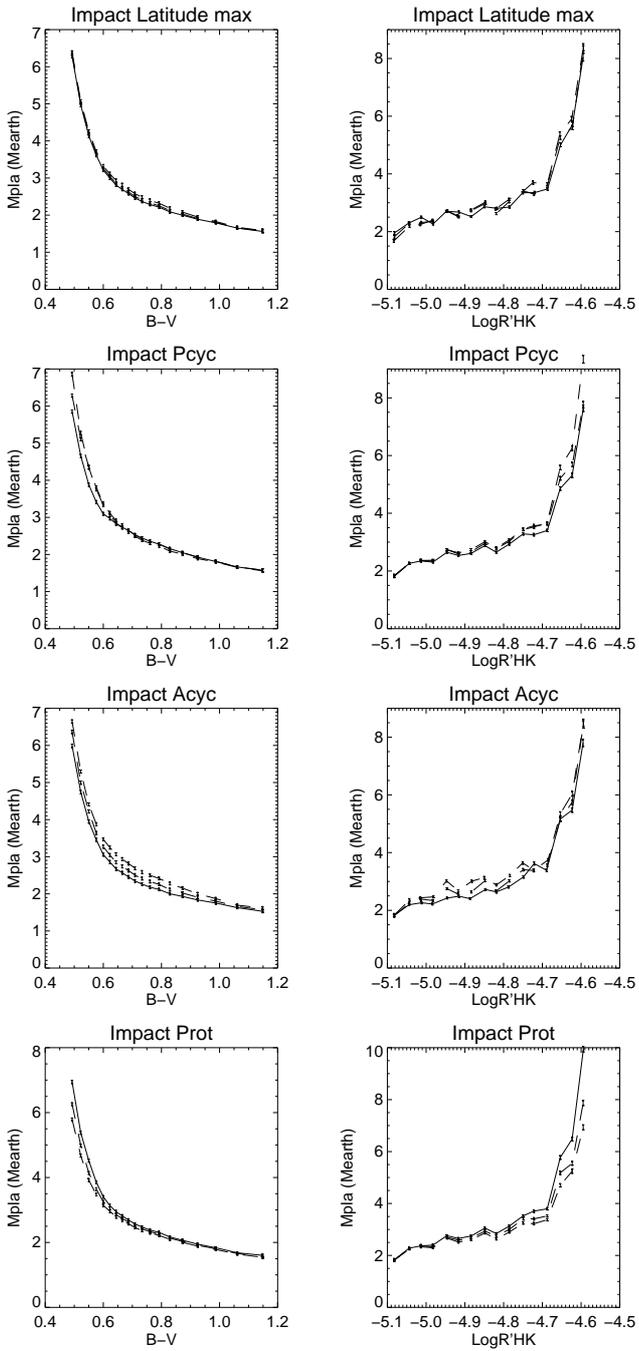}
\caption{
Effect of the parameters on M$_{\rm pla}$, similar to Fig.~4. 
}
\label{mpla_par}
\end{figure}

\begin{figure}
\includegraphics{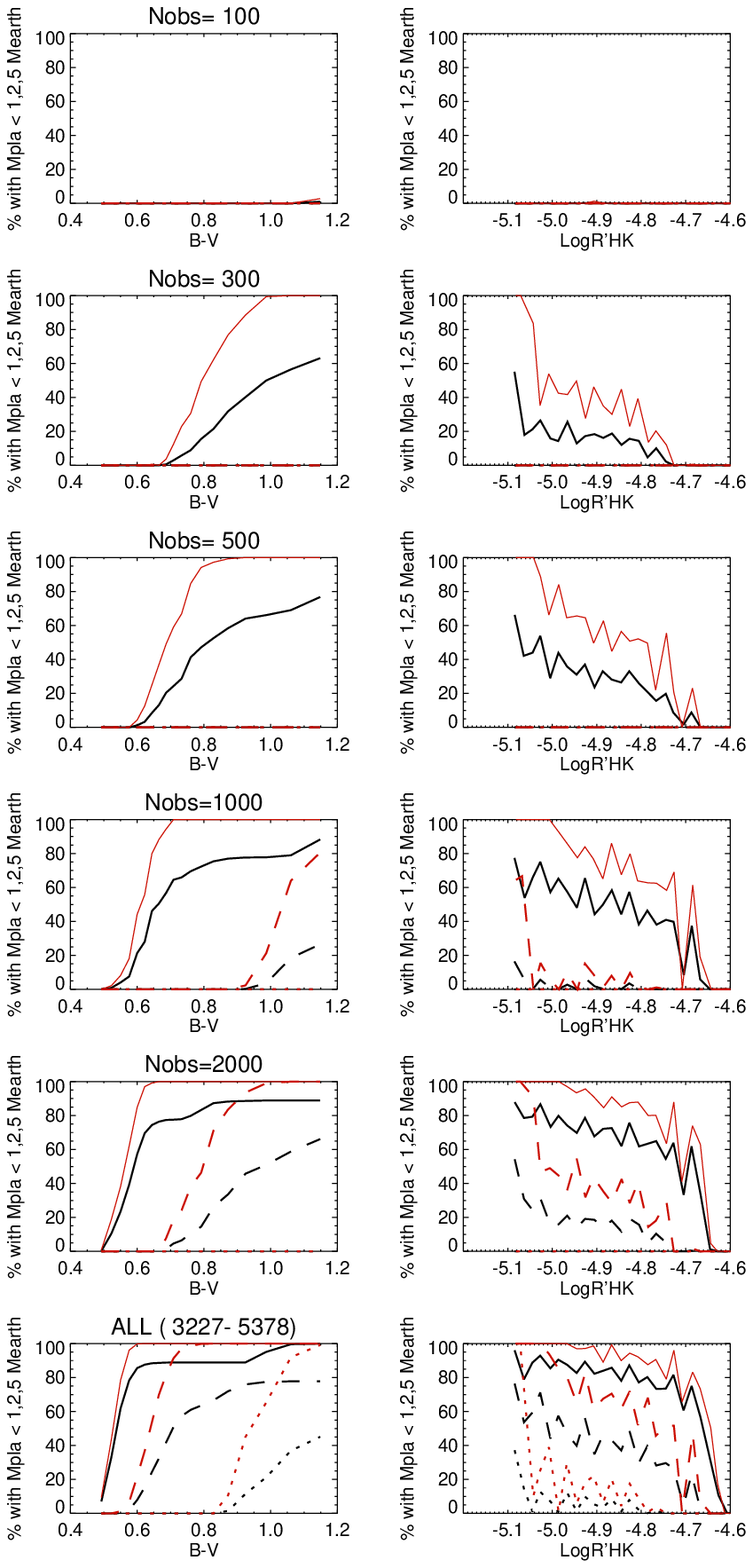}
\caption{
Percentage of simulations with M$_{\rm pla}$ below 5 M$_{\rm earth}$ (solid line), below 2 M$_{\rm earth}$ (dashed line), and 1 M$_{\rm earth}$ (dotted line), vs. B-V (left) and vs. LogR'$_{\rm HK}$ (right): for all inclinations except i=0 (black) and for edge-on configurations only (red). 
}
\label{mpla_pc}
\end{figure}

Fig.~\ref{kpl} shows $K_{\rm pl}$ (i.e., after correction using the correlation with LogR'$_{\rm HK}$ and a second-degree polynomial in time) versus B-V and LogR'$_{\rm HK}$ for various numbers of points and inclinations. 
The trends are naturally very similar to those of {\it rvact1} or {\it rvact2} shown in Fig.~\ref{composante}, but the amplitude was reduced compared to the original RV jitter due to the correction applied to the RV time series.
$K_{\rm pl}$ can be higher than a few m/s in the less favorable cases, but often reaches values below 1 m/s for stars with a sufficient number of observations, even for relatively active stars.

\subsection{Planetary mass and detection percentage}

From $K_{\rm pl}$, we derived a planet mass for a period that corresponds to the middle of the habitable zone and no eccentricity.  The habitable zone was  computed from the relationship between luminosity L and stellar flux S at the boundaries $\sqrt{L/S}$ \cite[][]{kasting93}, where the luminosity is derived from the effective temperature \cite[][]{zaninetti08}, as well as the flux at the boundaries \cite[][]{jones06}. This leads to a distance from the star of 1.73 AU (0.83 AU) and a period of 759 days (327 days) for F6 (K4) stars.   
We then divided the mass by sin(i), assuming that the planetary orbit lies within the equatorial plane of the star. Naturally, there may occasionally be departures from this configuration,
but very highly inclined orbits are unlikely. The RV signal due to activity indeed changes with inclination, as studied in the present paper, but the RV signal due to a planet will also change with inclination. This strongly affects the comparison with the RV jitter, which correspond to all types of inclinations.  


Fig.~\ref{mpla} show the resulting minimum planet mass vs. B-V and LogR'$_{\rm HK}$. Only a very large number of points (several thousands) allows reaching masses below 1 M$_{\rm earth}$ with the current status of correction techniques  (as defined by a criterion $C$=7.5, as discussed above), and this is only true for relatively quiet low-mass stars or stars with a medium activity level. 
A star like the Sun (G2, medium activity level) remains in a regime where the detection of planets with masses below 1~M$_{\rm earth}$ is expected to be very difficult, and possible only if there are a very large number of observations (several thousands) and if the star is seen close to edge-on. 
The lower limit can be much higher depending on the configuration. 
The mass is also very dependent on inclination because the RV jitter depends on inclination, and because low-inclination stars (assuming a planet orbiting in the equatorial plane) have planets with a lower RV amplitude. 
Fig.~\ref{mpla_par} shows the effect of the parameters on the mass. Overall, they have a relatively weak effect on the results, probably because they do not strongly affect  the RV jitter. The criterion used here, based on the RV jitter, is not necessarily the best metric to observe any effect because it does not take the temporal variability at different scales into account. More sophisticated methods must therefore be implemented to account for this.
The lower bound for the planet mass versus B-V for a number of points that are representative of well-observed stars (100-500, because there are very few stars with a larger number of nights) are in good agreement with the lower bound in a similar plot for exoplanets that were found using RV in the exoplanet encyclopedia (exoplanet.eu). 

Finally, Fig.~\ref{mpla_pc} shows the percentage of simulations providing masses below 1, 2, or 5 M$_{\rm earth}$. This quantifies the performance of current techniques in a simple way for various spectral types and activity levels. We first consider all inclinations. For low-mass stars (B-V above 0.9) and low-activity levels (LogR'$_{\rm HK}$ below -5.0), a small fraction (above 10\%) of planets with 1 M$_{\rm earth  }$ can be detected. For the same ranges, almost all 5 M$_{\rm earth}$ would be detected. However, the percentage falls to zero in the other ranges of B-V and LogR'$_{\rm HK}$. 
These percentages are only indicative and should be taken with caution because the same probability is attributed to each point in the grid in this computation.  

In the context of the follow-up programs of transit detections (TESS and PLATO) using RV techniques, it is interesting to consider simulations made edge-on only. The percentages, shown in red in Fig.~\ref{mpla_pc},  are higher and can be close to 100\% for the lowest mass star in the grid and a threshold of 1~M$_{\rm earth}$ (dashed lines). This percentage is different from 0 for B-V higher than 0.8 only, however. The edge-on configuration is then the most suitable inclination for detecting such planets, with good performance mostly for K stars if the sampling is very good.


\section{Conclusion}

We simulated RV time series showing variability for a range of spectral types from F6 to K4. These time series are realistic in the sense that the set of stellar parameters is consistent over this range of stars with different typical activity levels and variability. They also exhibit complex activity patterns,  which allowed us to retrieve a realistic complexity of the time series.
We here studied the effect of the parameters on the RV jitter. We showed that $\theta_{\rm max}$ and A$_{\rm cyc}$  strongly affect the RV jitter. We highlighted that the inclination of the star is a critical factor and that it leads to a wide diversity of configurations. The RV jitter is sensitive to the various contributions to the RV, but also to processes that operate at different timescales (we focused here on the rotation period and shorter periods on one hand, and on longer periods on the other): many configurations exist depending on these parameters. It is therefore crucial not to overinterpret observations because a local RV jitter is not entirely representative of the long-term variability, for example. 

The trend of the RV jitter with B-V and LogR'$_{\rm HK}$ is similar to the observed values \cite[][]{saar98,santos00,wright05}. However, the simulated RV jitter is lower. We estimate that such large jitter (e.g., between 5-10 m/s for stars with LogR'$_{\rm HK}$ below -5) cannot be reached with stellar activity (even when stellar granulation or supergranulation is included). The high observed RV jitter in the literature for these quiet stars may be due to either an underestimation of the instrumental noise, and/or the presence of other sources of RV (e.g., still-undetected planets). 
Finally, the slope of RV versus R'$_{\rm HK}$ obtained in our simulation is in good agreement with the observations \cite[][]{lovis11b}, both in range and trend versus spectral type, showing that the models providing RV and LogR'$_{\rm HK}$ give coherent results. They suggest, however, that the trend in the attenuation of the convective blueshift with T$_{\rm eff}$ observed by \cite{meunier17b} may be slightly stronger than estimated.

Using a simple approach based on the criterion proposed in \cite{dumusque17} and assuming a threshold that corresponds to the current performance of correction techniques, we showed that an Earth-mass planet in the middle of the habitable zone of these stars cannot be detected around the lower mass stars in our sample, and this is thought to be possible only with a very large number of observations. Better correction techniques must therefore be implemented in the future. New methods that have recently been proposed and are based on the fact that the convective blueshift depends on the line properties so that different sets of lines produce different RV time series or allow optimizing the RV \cite[][]{meunier17c,dumusque18} may be better adapted to the search for Earth-mass planets around solar-type stars.  In addition, further analysis of this large number of synthetic time series will greatly help us to find new paths to improve these correction methods. This will be done in \cite{meunier19c}.

\begin{acknowledgements}

This work has been funded by the ANR GIPSE ANR-14-CE33-0018.
We are very grateful to Charlotte Norris, who has provided us the plage contrasts we used in this work prior the publication of her thesis. 
This work was supported by the "Programme National de Physique Stellaire" (PNPS) of CNRS/INSU, which is cofunded by CEA and CNES.
This work was supported by the Programme National de Plan\'etologie (PNP) of CNRS/INSU, which is cofunded by CNES.

\end{acknowledgements}

\bibliographystyle{aa}
\bibliography{biblio}

%

\end{document}